\documentclass[fleqn,usenatbib]{mnras}

\usepackage{newtxtext,newtxmath}

\usepackage[T1]{fontenc}
\usepackage{ae,aecompl}

\usepackage{graphicx}	
\usepackage{amsmath}	
\usepackage[flushleft]{threeparttable}
\usepackage{comment}
\usepackage{multirow}
\usepackage{color}

\newcommand{\LxLb}{$L_{\rm X}/L_{\rm bol}$}
\newcommand{\LeLb}{$L_{\rm EUV}/L_{\rm bol}$}

\title[EUV irradiation of planetary atmospheres]{EUV irradiation of exoplanet atmospheres occurs on Gyr timescales}
\author[G.\ W.\ King \& P.\ J.\ Wheatley]{
George W. King$^{1,2,3}$\thanks{E-mail: kinggw@umich.edu}
and Peter J. Wheatley$^{1,2}$\thanks{E-mail: p.j.wheatley@warwick.ac.uk}
\\
$^1$Department of Physics, University of Warwick, Gibbet Hill Road, Coventry, CV4 7AL, UK\\
$^2$Centre for Exoplanets and Habitability, University of Warwick, Gibbet Hill Road, Coventry, CV4 7AL, UK\\
$^3$Department of Astronomy, University of Michigan, Ann Arbor, MI 48109, USA\\
}

\date{Accepted XXX. Received YYY; in original form ZZZ}

\pubyear{2020}

\begin{document}
\label{firstpage}
\pagerange{\pageref{firstpage}--\pageref{lastpage}}
\maketitle

\begin{abstract}
Exoplanet atmospheres are known to be vulnerable to mass loss through irradiation by stellar X-ray and extreme-ultraviolet emission. We investigate how this high-energy irradiation varies with time by combining 
an empirical relation describing stellar X-ray emission 
with a second relation describing the ratio of Solar X-ray to extreme-ultraviolet emission. In contrast to assumptions commonly made when modelling atmospheric escape,
we find that the decline in stellar extreme-ultraviolet emission is much slower than in X-rays, and that the total extreme-ultraviolet irradiation of planetary atmospheres is dominated by emission after the saturated phase of high energy emission (which lasts around 100\,Myr after the formation of the star). The extreme-ultraviolet spectrum also becomes much softer during this slow decline. Furthermore, we find that the total combined X-ray and extreme-ultraviolet emission of stars occurs mostly after the saturated phase. Our results suggest that models of atmospheric escape that focus on the saturated phase of high-energy emission are over-simplified, and when considering the evolution of planetary atmospheres it is necessary to follow extreme-ultraviolet driven escape 
on Gyr timescales.
This may make it more difficult to use stellar age to separate the effects of photoevaporation and core-powered mass-loss when considering the origin 
the planet
radius valley.
\end{abstract}




\begin{table*}
  \caption{Power law indices describing the decline in time of stellar X-ray and EUV luminosities as a function of $B-V$ colour. X-ray indices are taken from \citet{Jackson2012} and EUV indices are calculated using the empirical Solar relation from \citet{King2018}. The boundary between X-ray and EUV wavebands is taken to be 0.1\,keV. 
}
  \begin{center}
  \label{tab:powerlaws}
    \begin{tabular}{c c c c c c c c}
    \hline\
    Group & ($B-V$) colour bin & X-ray index & Full EUV index & Hard EUV index & Soft EUV index & X-ray ratio$^\dagger$ & EUV ratio$^\dagger$ \\
    && $\alpha$ & $\alpha(\gamma+1)$ & $\alpha(\gamma_{\rm hard}+1)$ & $\alpha(\gamma_{\rm soft}+1)$\\
    \hline
    1 & $0.290 \leq (B - V) < 0.450$ & $-1.22\pm0.30$ & $-0.70\pm0.22$ & $-0.80^{+0.21}_{-0.27}$ & $-0.29^{+0.21}_{-0.09}$ & $2.0\pm1.6$ & $3.4\pm1.6$\\
    2 & $0.450 \leq (B - V) < 0.565$ & $-1.24\pm0.19$ & $-0.71\pm0.17$ & $-0.81^{+0.15}_{-0.23}$ & $-0.30^{+0.20}_{-0.07}$ & $4.3\pm2.1$ & $5.6\pm1.8$\\
    3 & $0.565 \leq (B - V) < 0.675$ & $-1.13\pm0.13$ & $-0.65\pm0.15$ & $-0.74^{+0.11}_{-0.19}$ & $-0.27^{+0.18}_{-0.06}$ & $1.7\pm1.0$ & $3.2\pm1.1$\\
    4 & $0.675 \leq (B - V) < 0.790$ & $-1.28\pm0.17$ & $-0.74\pm0.17$ & $-0.84^{+0.14}_{-0.23}$ & $-0.31^{+0.21}_{-0.07}$ & $2.4\pm1.6$ & $3.8\pm1.6$\\
    5 & $0.790 \leq (B - V) < 0.935$ & $-1.40\pm0.11$ & $-0.81\pm0.17$ & $-0.92^{+0.12}_{-0.23}$ & $-0.33^{+0.23}_{-0.06}$ & $1.4\pm0.8$ & $2.8\pm1.1$\\
    6 & $0.935 \leq (B - V) < 1.275$ & $-1.09\pm0.28$ & $-0.63\pm0.20$ & $-0.71^{+0.20}_{-0.25}$ & $-0.26^{+0.19}_{-0.08}$ & $4.2\pm2.0$ & $5.5\pm1.8$\\
    7 & $1.275 \leq (B - V) < 1.410$ & $-1.18\pm0.31$ & $-0.68\pm0.22$ & $-0.77^{+0.22}_{-0.27}$ & $-0.28^{+0.20}_{-0.09}$ & $3.6\pm2.0$ & $4.9\pm1.8$\\
    \hline
\end{tabular}   
\end{center}
$^\dagger$ Ratio of the total integrated energy between 100\,Myr -- 1.0\,Gyr and 0 -- 100\,Myr
\end{table*}

\section{Introduction}
\label{sec-intro}
X-ray and extreme-ultraviolet (EUV; together, XUV) radiation is thought to be important in driving atmospheric escape from planets and exoplanets \citep[e.g.][]{Watson1981,Schneider1998,Lammer2003} and stellar X-ray luminosities are known to decline over a star's lifetime, as the star spins down due to magnetic braking \citep[as first suggested by][]{Schatzman1962}. For the first 100\,Myr or so, when the star is rapidly rotating, the X-ray emission appears independent of rotation period and is described by an approximately constant ratio of X-ray and bolometric luminosities, \LxLb$\approx10^{-3}$ \citep[e.g.][]{Vilhu1984,Vilhu1987,Wright11}. During this interval the high energy emission is considered to be {\em saturated}. 
This saturated interval is followed by a power law decrease in \LxLb, with a power-law index of around $-1.1$ to $-1.4$ \citep{Ribas2005,Jackson2012}. The intense X-ray emission during the saturated phase, followed by the steep decline, has led to the conclusion that the high-energy irradiation of a planetary atmosphere and the resulting atmospheric escape is dominated by the saturated phase 
\citep[e.g.][]{Lopez2013,Lammer2014,Owen2017}. 
Indeed, as pointed out by \citet{Owen2017}, a power-law decline significantly steeper than $-1$ would imply that the total high-energy emission of the star is dominated by the saturated phase. 

EUV fluxes for most stars are impossible to measure due to strong interstellar absorption in this waveband, and typically it is assumed that the 
EUV irradiation of planetary atmospheres declines at the same rate as in X-rays \citep[e.g.][]{Owen2012,Owen2013,Jin2014,Kurokawa2014,Luger2015,Ginzburg2016,Fossati2017,Owen2017,Fleming2019,Rogers2020}.
However, 
studies of high-energy Solar observations by \citet{Chadney2015} and \citet{King2018} have noted that Solar EUV emission remains relatively strong as X-ray surface flux decreases. 
Extrapolation of this Solar relation to higher activity levels successfully reproduces the observed X-ray to EUV flux ratios of nearby active stars \citep{Chadney2015,King2018} and the same relation predicts EUV fluxes of exoplanet host stars that are consistent with those extrapolated from Lyman-$\alpha$ observations \citep{Ehrenreich2015,Youngblood2016,Bourrier2017}. This empirical relation between X-ray and EUV fluxes implies that the EUV irradiation of exoplanet atmospheres may decline more slowly in time than X-rays. Consequently, significant EUV-driven atmospheric escape might persist well beyond the saturated phase of high-energy emission. 

On-going EUV-driven mass-loss
would be important in considering the evolutionary state of individual exoplanets, as well as the exoplanet population as a whole. Indeed, photoevaporation 
has been suggested to sculpt
both the Neptunian desert \citep{Mazeh2016,Owen2018} and the planet radius valley \citep{Fulton2017,Owen2017,VanEylen2018}.
Understanding the time dependence of photoevaporation may allow its effect to be separated from competing processes. The other leading proposed mechanism is core-powered mass-loss, in which atmospheric escape is driven by the gradual dissipation of the energy in the planetary core following its formation \citep{Ginzburg2016,Ginzburg2018,Gupta2019,Berger2020}. The timescale for the bulk of the escape is set by the cooling time of the core. Mass loss is predicted to occur on timescales of Gyr, and it has been suggested that this slower evolution might allow the effects of core-powered mass loss to be distinguished from photoevaporation \citep[e.g.][]{Gupta2019}.

In this letter, we combine empirical relations for the time evolution of stellar X-ray emission with relations for  X-ray to EUV flux ratios for the Sun in order to assess the time evolution of the EUV and total XUV irradiation of exoplanet atmospheres.

\section{Empirical stellar activity relations}
\subsection{X-ray flux evolution with time}
The evolution in time of average stellar X-ray fluxes is reasonably well defined, including for Sun-like stars by \citet{Ribas2005} and \citet{Claire2012}, and for a wider range of spectral types by \citet{Jackson2012}. \LxLb\ is seen to evolve with age, $t$, as
\begin{equation}
\frac{L_{\rm X}}{L_{\rm bol}} = \begin{cases}
\left(\frac{L_{\rm X}}{L_{\rm bol}}\right)_{\rm sat}, &\text{for } t < t_{\rm sat},\\
\left(\frac{L_{\rm X}}{L_{\rm bol}}\right)_{\rm sat} \times \left(\frac{t}{t_{\rm sat}}\right)^{\alpha}, &\text{for } t > t_{\rm sat},
\end{cases}
\label{eq:Jackson}
\end{equation}
where $\left(\frac{L_{\rm X}}{L_{\rm bol}}\right)_{\rm sat}$ is the value of \LxLb\ during the saturation interval, and $t_{\rm sat}$ is the length of that interval, typically around 100\,Myr. The power-law indices, $\alpha$, from \citet{Ribas2005} and \citet{Claire2012} were $-1.27$ and $-1.21\pm0.50$ respectively, and the values from
\citet{Jackson2012} are reproduced in Table~\ref{tab:powerlaws}. It is worth noting that most of these power-law indices are only marginally steeper than $-1$, implying that significant X-ray emission is to be expected beyond the saturated interval. 

The hard end of the EUV spectrum (100--360\,\AA) was also studied for nearby Sun-like stars by \citet{Ribas2005} and 
\citet{Claire2012}.
They found power-law indices of $-1.20$ and  $-1.09\pm0.48$ respectively, similar to their values for soft X-rays (20--100\,\AA). 
However, the uncertainties are large, and it was not possible to observe the softer end of the EUV (360--920\,\AA) due to interstellar absorption. 

\subsection{EUV flux evolution with X-ray flux}
\label{sec-euv}
The Sun is the only star for which we can study flux variations across the entire EUV spectrum. Analysis by \citet{Chadney2015} showed that the total EUV flux measured with the \textit{TIMED/SEE} instrument \citep{Woods2005} declines less steeply than X-ray flux as a function of the surface X-ray flux. They derived an empirical relation of the form
\begin{equation}
    \frac{F_{\rm EUV}}{F_{\rm X}} = \beta \left(F_{\rm X}\right)^\gamma,
\label{eq:Chadney}
\end{equation}
where $\beta$ and $\gamma$ are constants, and $F_{\rm EUV}$ and $F_{\rm X}$ are the EUV and X-ray surface fluxes respectively. Here, surface flux $F_{\rm EUV}$ = $L_{\rm EUV}/A$, where $A$ is the surface area of the star, with an equivalent definition for the X-ray band. \citet{Chadney2015} also demonstrated that this empirical relation could be extrapolated successfully to more active stars of different spectral types. 

In \citet{King2018}, we extended the analysis of \citet{Chadney2015} using additional \textit{TIMED/SEE} data, and by correcting an issue with the estimated uncertainties in some \textit{TIMED/SEE} data files. We also calculated values of $\beta$ and $\gamma$ for different choices of the boundary between X-ray and EUV wavebands in order to match the bandpasses of various X-ray telescopes. 
The power-law indices we obtained varied in $\gamma$ between $-0.425\pm0.110$ for a 0.1\,keV boundary (appropriate for \textit{ROSAT-PSPC}) and $-0.539\pm0.140$ for a 0.243\,keV boundary (appropriate for \textit{Chandra-ACIS}). 

For this letter, the uncertainties in the power-law indices have been estimated 
by dividing the \textit{TIMED/SEE} data into 24 equal time intervals (each spanning roughly six months) and fitting individually. 


\subsection{EUV flux evolution with time}
\label{sec-EUVvTime}

We can use this observed relationship between EUV and X-ray fluxes for the Sun (and verified by comparison with more active stars) to calculate the expected rate of EUV decline in time for the Sun and other stars by combining it with the X-ray time evolution determined by \citet{Jackson2012}. 
By substitution and rearranging Eqn.\,\ref{eq:Chadney},
\begin{equation}
\frac{L_{\rm EUV}}{L_{\rm bol}} = \beta \left( \frac{L_{\rm bol}}{A} \right)^{\gamma} \left( \frac{L_{\rm X}}{L_{\rm bol}} \right)^{\gamma + 1},
\end{equation}
where $A$ is the surface area of the star. Substituting for \LxLb\ from Eqn.\,\ref{eq:Jackson} for the unsaturated regime ($t > t_{\rm sat}$) gives
\begin{equation}
\frac{L_{\rm EUV}}{L_{\rm bol}} = \beta \left( \frac{L_{\rm bol}}{A} \right)^{\gamma} \left( \frac{L_{\rm X}}{L_{\rm  bol}}\right)^{\gamma + 1}_{\rm sat}  \left( \frac{1}{t_{\rm sat}}\right)^{\alpha (\gamma + 1)} t^{\alpha (\gamma + 1)}.
\label{eq:final}
\end{equation}

By way of example, taking a Solar-like star \citep[$\alpha = -1.13\pm0.13$ from][]{Jackson2012} and setting the boundary between X-ray and EUV wavebands at 0.1\,keV 
($\gamma=-0.425\pm0.110$ from Section\,\ref{sec-euv})
the EUV decline in time after the saturated period is described by a power law of index $-0.65\pm0.15$. We present EUV power law indices for all groups of stars from \citet{Jackson2012} in Table\,\ref{tab:powerlaws}.

\begin{figure}
\centering
 \includegraphics[width=\columnwidth]{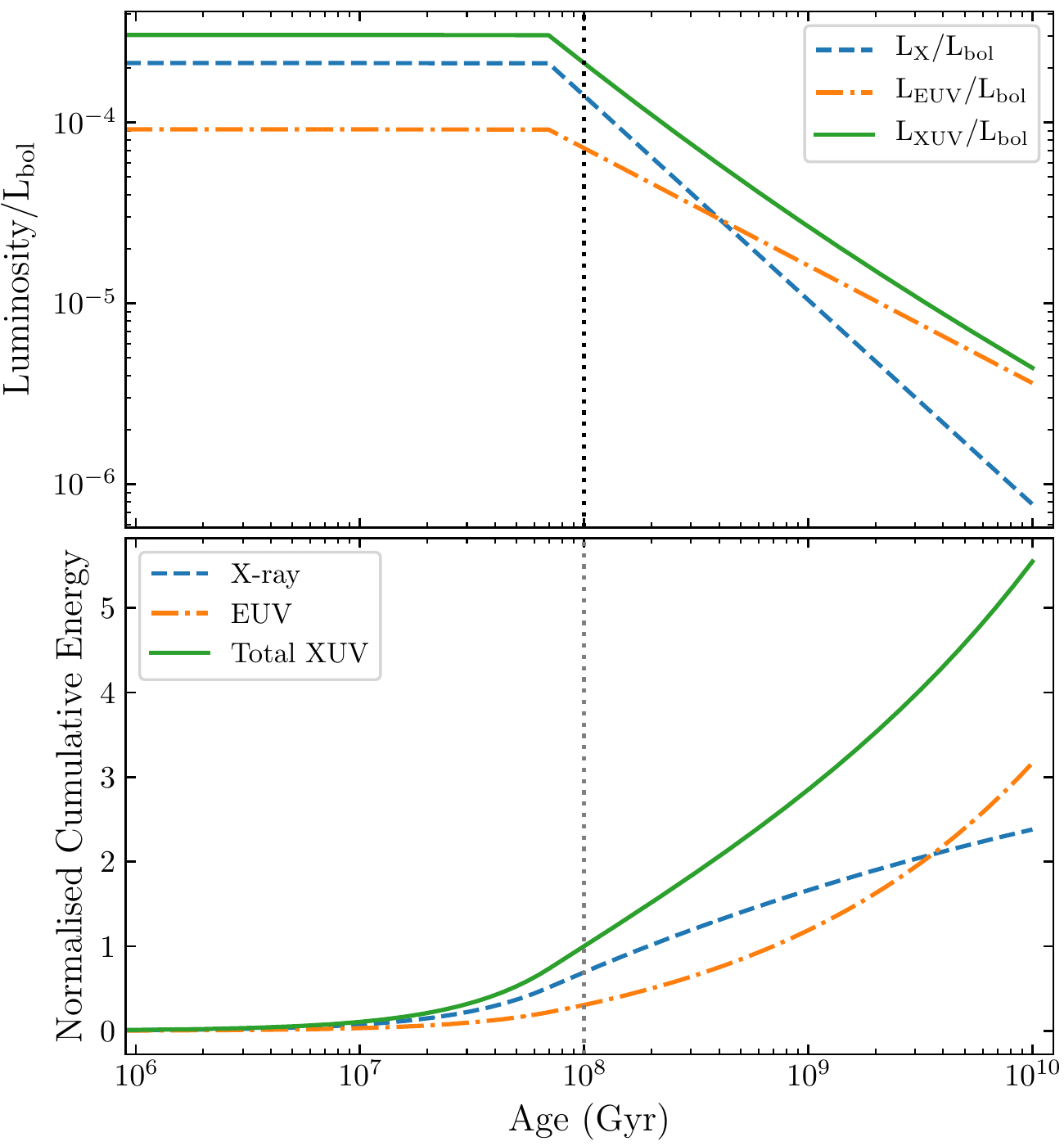}
 \caption{Top panel: Time evolution of the high-energy luminosity of a Sun-like star as determined in Section\,\ref{sec-EUVvTime}. X-ray luminosity is indicated by the dashed blue line, EUV luminosity by the dot-dashed orange line, and the total XUV luminosity by the solid green line. 
Bottom panel: Cumulative high-energy emission of a Sun-like star in same bands, normalised to the cumulative XUV emission at 100\,Myr. 
}
 \label{fig:log}
\end{figure}

In the upper panel of Fig.~\ref{fig:log}, we show the time evolution of \LxLb\ and \LeLb\ for a Solar-like star, illustrating the slower decline of EUV emission after the saturated interval. The lower panel of Fig.~\ref{fig:log} shows the cumulative energy output in each band, and the total across both. The cumulative plot shows that while the X-ray energy is seen to gradually level off, the corresponding EUV energy continues to increase sharply, even at late times. Only around 10 per cent of the lifetime EUV energy, and 20 per cent of the total XUV energy, is emitted during the first 100\,Myr. This suggests that significant EUV-driven atmospheric escape from exoplanets may persist for Gyr timescales, which is much longer than assumed in the theoretical studies highlighted in Section\,\ref{sec-intro}.

It is also worth noting here that there is as much X-ray energy emitted between 100\,Myr and 1\,Gyr as there is in the first 100\,Myr. Two-thirds of the lifetime X-ray emission occurs after 100\,Myr. 

Table\,\ref{tab:powerlaws} shows the power-law index of the EUV decline is shallower than $-1$ for the full range of FGK spectral types studied by \citet{Jackson2012}. Therefore, in all cases we can expect the majority of EUV energy to be emitted 
on Gyr timescales.
Table\,\ref{tab:powerlaws} shows that FGK stars typically emit 3--6 times more EUV radiation between the age of 100\,Myr and 1\,Gyr than they do during their first 100\,Myr. 

\subsection{The softer the photons, the slower the decline}
\label{sec:softest}

The rates of decline we find here for the entire EUV band (0.0136--0.1\,keV; 124--912\,\AA) are shallower than those found by \citet{Ribas2005} and \citet{Claire2012} for the harder end of the EUV (100--360\,\AA). They are also shallower than the slope inferred from coronal models by \citet{Sanz-Forcada2011}.
A possible explanation for the different slopes is that the rate of EUV decline varies between the observable hard end of the EUV spectrum and the unobservable soft band (which is obscured by interstellar absorption even for nearby stars). To investigate this possibility, we reanalysed the Solar \textit{TIMED/SEE} data, separating the hard and soft EUV bands. 

\begin{figure}
\centering
 \includegraphics[width=0.88\columnwidth]{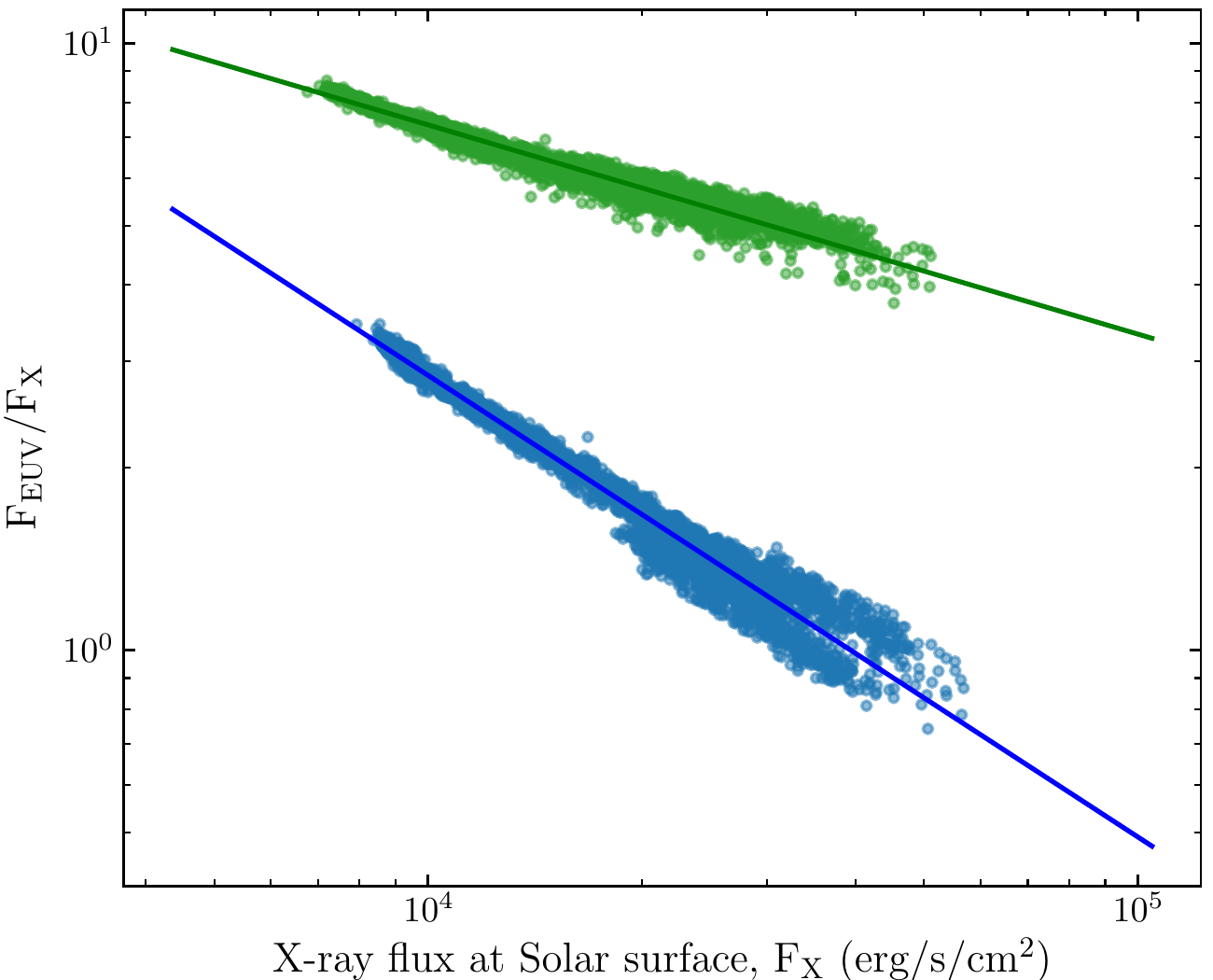}
 \caption{The ratio of Solar EUV to X-ray flux as a function of surface X-ray flux for two different EUV bands, measured using the \textit{TIMED/SEE} instrument. The green data points and power-law fit are for the hard end of the EUV spectrum (100--360\,\AA), which is observable for nearby stars, while the blue points and power law represent the soft EUV (360--920\,\AA), which suffers very strong interstellar absorption even for the nearest stars. It is clear that these two EUV bands have different power-law dependence on X-ray flux. 
 }
 \label{fig:hardsoft}
\end{figure}

Fig.~\ref{fig:hardsoft} shows the ratio of Solar EUV to X-ray flux, as a function of X-ray surface flux, for both the hard and soft EUV bands as defined by \citet{Ribas2005} and \citet{Claire2012} (100--360\,\AA, and 360--920\,\AA). It can be seen that the power-law index is indeed markedly different for these two EUV bands, with $\gamma_{\rm hard}=-0.35^{+0.07}_{-0.15}$ and $\gamma_{\rm soft}=-0.76^{+0.16}_{-0.04}$. The associated values of $\beta$ are 116 and 3040\,erg\,s$^{-1}$\,cm$^{-2}$, respectively. In Table\,\ref{tab:powerlaws}, we give the implied EUV decline in time for each EUV band and spectral type using the X-ray decline rates from \citet{Jackson2012}.
For a Solar-type star, we find a time evolution power-law index of $-0.74^{+0.11}_{-0.19}$ for the hard EUV, which is consistent with \citet{Ribas2005} and \citet{Claire2012}, and a
much shallower slope of $-0.27^{+0.18}_{-0.06}$ for the soft end of the EUV spectrum. This is characteristic of chromospheric activity decline determined by \citet{Claire2012}.

\section{Discussion}
A useful first approximation for the mass loss rate from an exoplanet atmosphere is energy-limited escape \citep{Watson1981,Erkaev2007}. Here one considers the atmospheric escape to be directly proportional to the input XUV energy, with that energy overcoming gravitational potential energy at some efficiency, typically assumed to be about 15 per cent \citep[e.g.][]{Watson1981,Tian2005,Kubyshkina2018}. As outlined in Section\,\ref{sec-intro}, it is often assumed that the XUV input to the exoplanet atmosphere, and hence its mass loss, is dominated by the saturated phase of stellar activity during the first 100\,Myr or so of the lifetime of the star. 

The slow decline of EUV emission we infer in Section\,\ref{sec-EUVvTime} suggests that most high-energy irradiation of exoplanet atmospheres occurs on Gyr timescales, continuing well after the initial 100\,Myr (see Fig.\,\ref{fig:log}). We should therefore expect significant evolution of exoplanet atmospheres well beyond the saturated phase of stellar activity. In Table\,\ref{tab:powerlaws} we show that the EUV emission between 100\,Myr and 1\,Gyr exceeds the emission in the first 100\,Myr by factors of 3--6 across spectral types FGK. Furthermore, as can be seen in Fig.\,\ref{fig:log}, the integrated EUV emission continues to increase significantly well beyond 1\,Gyr. 


\citet{MurrayClay2009} modelled a hydrodynamic planetary wind from a hot Jupiter and found that simple energy-limited escape is a good approximation when irradiating XUV fluxes are low. 
For the higher fluxes at earlier times, a substantial proportion of the input energy is lost to radiative cooling, 
and atmospheric escape is less efficient. 
Radiative cooling in this "recombination-limited" regime of atmospheric escape therefore further weights mass loss to late times (i.e.\ even more than the integrated energy flux shown in Fig.\,\ref{fig:log}). \citet{Owen2012} also describes how the outflows of atmospheric material are EUV-driven at lower X-ray fluxes, after a few hundred Myr, even when considering the EUV time evolution as equivalent to the X-ray. For smaller planets, the numerical calculations of atmospheric escape from Neptune-mass planets by \citet{Ionov2018} also highlight how energy-limited estimates are a good approximation when irradiating fluxes are low, and that mass loss efficiency is lower for higher irradiating fluxes at earlier times (see their Fig.\,3).


\citet{Owen2016} identified a third regime of "photon-limited" atmospheric escape, where the mass loss rate is set by the flux of incoming ionising photons, and which applies at low irradiating fluxes for planets with shallow gravitational potentials. The much slower rate of EUV decline we find for the softest ionising photons (see Section\,\ref{sec:softest}) implies that photon-limited escape is even further weighted to late times, since the average energy of ionising photons continues to decrease and hence the energetic efficiency of mass loss will increase. This softening of the EUV spectrum therefore acts to extend atmospheric escape to even later times for low mass planets, with the unobservable soft end of the EUV likely to 
dominate 
at late times.


The softening of the EUV spectrum suggests that as the magnetic dynamos of late-type stars weaken with age, this is accompanied by ever lower average plasma temperatures in the corona. The softening of stellar spectra with age, characteristic of lower coronal plasma temperatures, is well known from observations at X-ray wavelengths across different stellar ages \citep[e.g.][]{Favata2003,Telleschi2005} and our results show this extends further into the EUV at late times and low activity levels. The shallow decline we see for the soft end of the EUV is also similar to slopes observed for chromospheric emission at longer wavelengths in the FUV \citep[e.g.][]{Claire2012}.

The correct time-dependence of EUV and XUV irradiation is necessary for understanding the evolution of the exoplanet population as a whole. An important example is the radius valley found between super-Earths and sub-Neptunes with \textit{Kepler} \citep{Fulton2017,VanEylen2018}. This had been predicted to arise due to atmospheric escape of primordial H/He envelopes driven by intense XUV irradiation during the saturated phase of stellar activity \citep{Owen2013,Lopez2013,Owen2017}.  However, as an alternative explanation, \citet{Ginzburg2018} and \citet{Gupta2019} have shown that the luminosity of the cooling planetary core is also sufficient to strip primordial H/He envelopes, and that this process also leads naturally to a bimodal distribution of planet radii. One way to unravel the relative importance of XUV irradiation and core cooling in sculpting the radius valley
is to consider the time-dependence of atmospheric escape, i.e.\ testing mass-loss predictions against populations of planets selected by age \citep{Berger2020}. Core-powered mass loss will occur on timescales of Gyr \citep{Ginzburg2018,Gupta2019}, whereas it has been expected that photoevaporation will be dominated by the first 100\,Myr \citep{Lopez2013,Lammer2014,Owen2017,Rogers2020}. Our results suggest that photoevaporation may act on longer timescales than has generally been considered. 

In order to assess the timescale on which XUV-driven mass loss proceeds, it is also necessary to consider how radius of the planet evolves with time. In general, mass loss and the cooling of the envelope leads to a decreasing radius with time, which acts to reduce the mass loss rate \citep[e.g.][]{Owen2013,Lopez2013,Lopez2017,Kubyshkina2020}. This will tend to weight atmospheric escape to earlier times, and so it remains possible that XUV-driven mass loss occurs primarily during the saturated phase of XUV emission, within the first 100 Myr.
Nevertheless, our results showing that the total integrated EUV fluxes are dominated by the timescales of Gyr must be considered and included in future models of atmospheric escape.

In this work, we have assumed the X-ray-age relations of \citet{Jackson2012} as a basis to consider how the EUV and total XUV fluxes evolve with time. 
However, there are indications that X-ray evolution in time may be more complex than found by \citet{Jackson2012}. For example, \citet{Tu2015} suggested that a range of initial stellar spin periods leads to some stars spending much longer in the saturated phase than others (a range of 20--500\,Myr, instead of the canonical 100\,Myr). This range of stellar spin histories will lead to greater scatter in the integrated X-ray, EUV and total XUV emission plotted in Fig.\,\ref{fig:log}. However, our conclusion that the slow decline of EUV flux leads to late-time evolution of exoplanet atmospheres is unaffected. Indeed, for stars that form with relatively slow initial rotation, the integrated EUV emission after the saturated phase will be an even greater proportion of the lifetime XUV emission. 

For older stars, \citet{Booth2017} found evidence that the X-ray power-law decline may steepen beyond an age of around 1\,Gyr. This might reduce the importance of atmospheric escape at later times.  However, as the authors state, confirmation of this result requires further work with a larger sample of stars. Even if this late-time steep decline is confirmed, our conclusion that the EUV emission between 0.1--1\,Gyr greatly exceeds the saturated phase is unaffected (see Table\,\ref{tab:powerlaws}).

\section{Conclusions}
In this letter we have considered how stellar EUV emission varies with stellar age. By combining empirical relations for stellar X-ray emission with empirical relations for Solar EUV to X-ray ratios, we find evidence that stellar EUV emission declines in time with power-law indices substantially shallower than $-1$ (see Table\,\ref{tab:powerlaws}). We also find that soft end of the EUV band the declines much more slowly than the harder EUV.

Our results imply that significant EUV irradiation of exoplanet atmospheres continues on Gyr timescales, and that the resulting photoevaporation may not be dominated by the saturated activity phase during the first 100\,Myr (see Fig.\,\ref{fig:log}). For instance, we find that EUV emission in the interval 0.1--1\,Gyr exceeds emission during the first 100\,Myr by a factor 3--6 (see Table\,\ref{tab:powerlaws}).

We note that total EUV-driven mass-loss is weighted further to late times by the known decrease in photoevaporation efficiency with irradiating flux (in the recombination-driven regime). In addition, 
the softening of the EUV spectrum during the decline increases the flux of ionising photons for a given energy flux. This weights EUV-driven mass loss to even later times in the photon-limited regime for planets with shallow gravitational potentials. 

While our results rely on the assumption that spectral variations observed for the Solar corona are characteristic of the long-term spectral evolution of stellar coronae, our approach has the advantage of being entirely empirical, with no need to employ poorly-constrained coronal models or to attempt to extrapolate X-ray spectra to the EUV waveband. Previous work has also shown that this approach successfully reproduces X-ray to EUV ratios of nearby active stars \citep{Chadney2015,King2018} and that our EUV estimates are consistent with extrapolations from Lyman-$\alpha$ \citep{Ehrenreich2015,Youngblood2016,Bourrier2017}. 

Our results strongly suggest that models of atmospheric escape need to account for EUV irradiation that is dominated by Gyr timescales, and not just the first 100\,Myr of the life of the system. This will be important in unravelling the relative roles played by photoevaporation and core-powered mass-loss in sculpting features of the exoplanet population such as the radius valley.


\section*{Acknowledgements}
The work presented here was supported by STFC consolidated grants ST/P000495/1 and ST/T000406/1.

\section*{Data Availability}
The \textit{TIMED/SEE} data used in this work are publicly available at \url{https://lasp.colorado.edu/home/see/data/daily-averages/level-3/}



\bibliographystyle{mnras}
\bibliography{euv_final} 








\bsp	
\label{lastpage}
\end{document}